\documentclass[preprint,12pt]{aastex}

\usepackage{latexsym}
\newcommand {\ha}{H$\alpha$}

\shorttitle{X-Ray--Bright Superbubbles in the LMC}
\shortauthors{Dunne, Chu, \& Points}

\begin{document}

\title{X-Rays from Superbubbles in the Large Magellanic Cloud. VI. 
A Sample of Thirteen Superbubbles}

\author{Bryan C. Dunne, Sean D. Points, and You-Hua Chu}
\affil{Department of Astronomy, 
University of Illinois,
1002 West Green Street, 
Urbana, IL 61801}
\email{carolan@astro.uiuc.edu, points@astro.uiuc.edu, chu@astro.uiuc.edu}

\begin{abstract}
We present {\it ROSAT} observations and analysis of thirteen 
superbubbles in the Large Magellanic Cloud.  Eleven of these 
observations have not been previously reported.  We have studied the 
X-ray morphology of the superbubbles, and have extracted and 
analyzed their X-ray spectra.  Diffuse X-ray emission is detected from 
each of these superbubbles, and X-ray emission is brighter than is 
theoretically expected for a wind-blown bubble, suggesting that the 
X-ray emission from the superbubbles has been enhanced by interactions 
between the superbubble shell and interior SNRs.  We have also found 
significant positive correlations between the X-ray luminosity of a 
superbubble and its \ha\ luminosity, expansion velocity, and OB star 
count.  Further, we have found that a large fraction of the superbubbles 
in the sample show evidence of ``breakout'' regions, where hot X-ray
emitting gas extends beyond the \ha\ shell.  

\end{abstract}


\clearpage
\section{Introduction}

Superbubbles are large ($\sim$100 pc across) shells in the
interstellar medium (ISM) created by the combined action of stellar
winds and supernova explosions of massive stars in an OB association.
The hot ($\gtrsim$10$^6$ K) shock-heated gas interior to superbubbles
emits X-ray radiation.  X-ray observations of superbubbles can reveal
a wealth of information on the structure and interior of superbubbles.
An excess of diffuse X-ray emission in superbubbles can indicate the
presence of interior supernova remnants (SNRs) shocking the inner
walls of the superbubble shell (\citealp{Paper1}, hereafter Paper I;
\citealp{Wang91}).  The diffuse X-ray emission can also be used to
find ``breakout'' regions where the hot gas from the superbubble
interior may be leaking out into the ISM.  Unresolved peaks superposed
on the diffuse X-ray emission may also indicate the presence of
stellar X-ray sources interior to the superbubble.

The Large Magellanic Cloud (LMC) provides an ideal laboratory for
observing superbubbles in the X-ray spectrum.  The coverage of
Galactic superbubbles is far from complete because of extinction from
the disk of the Milky Way.  Non-Magellanic Cloud extragalactic
superbubbles are too far away to be angularly resolved by X-ray
instruments such as {\it ROSAT}. The LMC, however, provides a sample
in excess of 20 superbubbles, at a common distance
\citep[$\sim$50~kpc,][]{Feast99}, that are resolvable by modern X-ray
detectors.  Observations of the superbubbles in the LMC can provide us
with great insight into the interaction among superbubbles, SNRs, and
the ISM.

We have been studying X-ray emission from superbubbles in the LMC.  In
Paper I, {\it Einstein} observations were used to show that seven LMC
superbubbles are diffuse X-ray sources with luminosities much higher
than those expected by the wind-blown, pressure-driven bubble models
of \citet{Weaver77}.  Off-center SNRs are proposed to be responsible
for the excess X-ray emission. {\it ROSAT} observations of the
superbubble N44 confirmed its diffuse X-ray emission and provided the
first useful X-ray spectra of N44 for determinations of plasma
temperatures \citep[hereafter Paper II]{Paper2}.  To illustrate that
excess X-ray emission from superbubbles is caused by an intermittent
process, \citet[hereafter Paper III]{Paper3} analyzed {\it ROSAT}
observations of four X-ray-dim superbubbles, and showed that these
superbubbles do not have excess X-ray emission.  For high-resolution
spectral analysis, {\it ASCA} observations of N44 were made; the {\it
ASCA} data showed that the hot gas in the breakout region is slightly
cooler than that in the superbubble interior \citep[hereafter Paper
IV]{Paper4}.  {\it ROSAT} observations of the \ion{H}{2} complex N11
were analyzed to study the interaction between OB associations,
\ion{H}{2} regions, and superbubbles \citep[hereafter Paper
V]{Paper5}.

In this latest study, we have analyzed {\it ROSAT} observations of
eleven H$\alpha$-indentified superbubbles in the LMC whose
observations had not been reported previously.  Diffuse X-ray emission
was detected in every one of these eleven superbubbles.  We have
re-analyzed N44 and N11 in order to have a homogeneous set of results
for comparisons.  We have modeled these superbubbles using the
pressure-driven bubble models of \citet{Weaver77}.  In this paper, we
report the {\it ROSAT} observations of the thirteen superbubbles
studied, and discuss the X-ray luminosities and other properties of
the superbubbles and their relationship with the LMC.

\section{X-Ray Dataset and Analysis}
\label{sec:data}

\subsection{{\it ROSAT} Archival Dataset}
Our dataset is based on a selection of known LMC superbubbles around
OB associations with a well-defined \ha\ morphology.  We have further
constrained the sample to include only those superbubbles with
previously unreported {\it ROSAT} observations with at least 5~ks
exposure.  The superbubbles studied are in N51, N57, N103, N105, N144,
N154, N158, N160, N206 \citep[nomenclature of][]{Henize56}, and
30~Dor~C.  Two superbubbles are present in N51, making the total
number of superbubbles eleven.  For comparisons with previous results,
we have also included N11 and N44 in the dataset.  The coordinates,
sizes, \ha\ luminosities, expansion velocities, and local OB
associations of this sample of thirteen superbubbles, as well as
alternative designations, are summarized in Table~\ref{tbl:Prop}.

Two detectors are available on board the {\it ROSAT} satellite: the
Position-Sensitive Proportional Counter (PSPC) and the High-Resolution
Imager (HRI).  We have used PSPC observations to investigate physical
conditions and distribution of the 10$^6$~K gas interior to most of
the LMC superbubbles in our dataset.  The PSPC is sensitive to X-ray
photons with energies in the range of 0.1--2.4 keV and has an energy
spectral resolution of $\sim$40\% at 1~keV, with a field of view of
$\sim$2$^{\circ}$.  As the HRI is better suited to revealing points
sources rather than diffuse emission, we have used HRI observations to
investigate the distribution of X-ray emitting gas only in the
superbubble N206, which fell close to the outer edge of the
field-of-view in the PSPC observations.  The HRI is sensitive in the
energy range of 0.1--2.0 keV, with a field of view of $\sim$40$'$.
Further information on the PSPC and HRI can be found in \citet{Pfef87}
and the \citet{ROSAT91}.  A summary of the individual observations can
be found in Table~\ref{tbl:PSPCobs}.

\subsection{X-Ray Data Analysis}
We have studied both the X-ray morphology and X-ray spectra of the
superbubbles in the dataset.  All of the data were reduced using
standard routines in IRAF\footnote{Image Analysis and Reduction
Facility -- IRAF is distributed by the National Optical Astronomy
Observatories, which are operated by the Association of Universities
for Research in Astronomy, Inc., under cooperative agreement with the
National Science Foundation.}  and the PROS\footnote{PROS/XRAY Data
Analysis System -- http://hea-www.harvard.edu/PROS/pros.html} package.

\subsubsection{Morphology}
Analysis of the X-ray morphology of each superbubble was conducted
using smoothed PSPC and HRI images.  The images were binned to
4\arcsec\ pixels and then smoothed with Gaussian function of $\sigma
=$ 4 pixels.  We have compared the X-ray morphologies with the \ha\
morphologies observed in the PDS scans of the Curtis Schmidt plates of
\citet{KH86}.  In Figure~\ref{fig:contour}, we present the \ha\ images
overlaid with X-ray contours.  We also present the X-ray images
overlaid with the same contours to ensure the clarity of the contour
levels.  The contours are at levels of 50\%, 70\%, 85\%, and 95\% of
the peak level within the superbubble.  For bright X-ray objects in
the field not actually part of the superbubble (such as SNRs), we have
plotted additional contours at 2, 4, 8, and 16 times the superbubble
peak level.  These contours are plotted as dashed lines.

\subsubsection{Spectral Fits}
The X-ray spectra of the superbubbles were extracted from the PSPC
data.  We defined source regions for each superbubble.  Then, possible
stellar sources (i.e., X-ray binaries, Wolf-Rayet stars) were excised
from the data before the spectra were extracted.  Additionally, we
selected several background regions around each superbubble (This is
especially important for superbubbles superposed on large extended LMC
sources of diffuse X-ray emission, such as the 30 Doradus complex and the
supergiant shells LMC2 and LMC3).  The background-subtracted spectra
were then extracted from the PSPC event files.

The observed X-ray spectrum of each superbubble is a convolution of
several factors: the intrinsic X-ray spectrum of the superbubble, the
intervening interstellar absorption, and the PSPC response function.
Because the interstellar absorption and the PSPC response function are
dependent on photon energy, we must assume models of the intrinsic
X-ray spectrum and the interstellar absorption to make the problem
tractable.  As the X-ray emission from the superbubble interiors
appears largely diffuse, we have used the \citet{RS77} thin-plasma
emission model and the \citet{Morr83} effective absorption
cross-section per hydrogen atom to describe the intrinsic X-ray
spectra of the superbubbles and the foreground absorption,
respectively.  We then simulate the observed spectrum, combining the
assumed models for the intrinsic spectrum and the interstellar
absorption with the response function of the PSPC.  The observed
spectrum is fitted with the simulated spectra; the $\chi^{2}$ of the
fits determines the best-fit.

We performed a $\chi^{2}$ grid search of simulated spectral fits to
determine the best-fit levels for the temperature, $kT$, and
absorption column density, $N_{\rm H}$. From these model fits, we can
calculate the un-absorbed X-ray flux, and therefore the X-ray
luminosity, $L_{\rm X}$, of the diffuse X-ray emission from the
superbubbles.  The normalization factor of the best fit can be used to
determine the volume emission measure, adopting a distance of 50~kpc
for the LMC \citep{Feast99}.  If we assume a uniform density in the
X-ray-emitting gas, the volume emission measure can be expressed as
N$_{\rm e}^{2}fV$, where N$_{\rm e}$ is the electron density, $f$ is
the volume filling factor, and $V$ is the volume of the superbubble
interior.  Assuming the superbubbles have an ellipsoidal shape, we can
use the observed diameters of the superbubbles to determine N$_{\rm e}
\sqrt{f}$.  The net exposure time, background-subtracted source
counts, scaled background counts, and best-fit values of $kT$, $N_{\rm
H}$, $L_{\rm X}$ and N$_{\rm e} \sqrt{f}$ for a 30\% solar abundance
\citet{RS77} thermal plasma model are given in
Table~\ref{tbl:FreeSpec}.  Plots of the fits to the superbubbles'
X-ray spectra are shown in Figure~\ref{fig:freespec}.

We have also sought to further constrain the models by using
observations of the \ion{H}{1} column density to independently
determine the total absorption column density, $N_{\rm H}$.  We have
divided the absorption column into Galactic and LMC components and
determined each separately.  \citet{AB99} demonstrated that in the
X-ray spectrum, at Galactic latitudes $|b| > 25^{\circ}$, the
contribution of molecular gas to the total absorption column density
is comparable to the contribution of the neutral hydrogen gas.  We
have therefore approximated the Galactic absorption column density by
$N_{\rm H} \simeq 2 \times N_{\rm HI}$.  Continuing this approximation
to the LMC absorption column density is more complicated.
Measurements of LMC \ion{H}{1} column density can sample material both
in front of and behind a feature such as a superbubble.  We have used
the simplifying assumption that half of the LMC \ion{H}{1} gas is
foreground to the superbubbles and half is background.  Thus, the LMC
component of the total absorption column density is $\frac{1}{2}
\times 2 \times N_{\rm HI} = N_{\rm HI}$.  The total absorption column
density is therefore given by $N_{\rm H} = 2 \times (N_{\rm HI})_{\rm
Galactic} + (N_{\rm HI})_{\rm LMC}$.

We have used the observations of Galactic and LMC \ion{H}{1} column
densities by \citet{DL90} and \citet{Rohlfs84} to determine the total
absorption column density.  These calculated values of $N_{\rm H}$
are, on average, nearly an order of magnitude larger than values of
$N_{\rm H}$ determined by the best-fit models to the X-ray spectra
(see Table~\ref{tbl:FixSpec}).  Indeed, several of the Galactic
$N_{\rm H}$ values are alone larger than those derived from the
best-fit models.  The calculated values of $N_{\rm H}$ also show a
much narrower range of absorption column densities to the LMC
superbubbles (log $N_{\rm H} =$ 21.3--21.7) than the values of $N_{\rm
H}$ derived from the best-fit models to the X-ray spectra (log $N_{\rm
H} =$ 20.4--22.0).

We have re-determined best-fit values for $kT$, $L_{\rm X}$, and
N$_{\rm e}$ based again on a 30\% solar abundance \citet{RS77} thermal
plasma model, but with $N_{\rm H}$ fixed at the calculated values
determined above.  A summary of the results are given in
Table~\ref{tbl:FixSpec}.  Plots of these ``fixed $N_{\rm H}$'' fits to
the superbubbles' X-ray spectra are shown in Figure~\ref{fig:fixspec}.

\section{Individual Superbubble Properties}
\label{sec:properties}
We will now discuss each of the superbubbles studied individually.
For each superbubble, although we primarily use the nomenclature of
\citet[e.g., N11]{Henize56}, we give alternative designations as
cataloged by \citet[e.g., DEM L 192]{DEM76}.  We also give the OB
associations and star clusters encompassed by the superbubble as
reported in \citet[e.g., LH63]{LH70}.  General descriptions of the
\ha\ morphology, as seen in the PDS scans of the Curtis Schmidt plates
of \citet{KH86}, and comparisons with X-ray morphology, from the
smoothed PSPC images, are given.  We also discuss breakout regions,
possible identifications of X-ray hotspots with known stellar sources
from \citet[e.g., Br81]{Brey81} and \citet[e.g.,
Sk$-$66$^{\circ}$28]{Sk69}, and other interesting X-ray features.  The
linear sizes of the superbubbles are calculated assuming 1$'$ $=$
15~pc.

\subsection{N11--Shell 1}
\label{sec:n11}
N11 (DEM~L~34) is the second largest \ion{H}{2} complex in the LMC
\citep{KH86}.  N11 contains a superbubble 150~pc $\times$ 100~pc in
size, surrounded by several bright \ha\ knots and smaller \ha\ shells.
The superbubble has been labeled Shell 1 in \citeauthor*{Paper5}.
Shell 1 encompasses the OB association LH~9.  Diffuse X-ray emission
is detected toward Shell 1 (See Figure~\ref{fig:contour}a).

This diffuse X-ray is centrally bright and appears to be confined by
the observed \ha\ shell.  The X-ray emission peaks at a location
coincident with HD~32228, a known Wolf-Rayet star also cataloged as
Br81 and Sk$-$66$^{\circ}$28.  Several smaller peaks are also evident
in the diffuse X-ray emission.  A more thorough interpretation of the
X-ray emission can be found in \citeauthor*{Paper5}.

\subsection{N44 (DEM~L~152)}
\label{sec:n44}
N44 is a bright \ion{H}{2} complex, similar to N11.  N44 contains a
superbubble, cataloged as DEM~L~152, around the OB association LH~47.
The 100~pc $\times$ 75~pc superbubble is well-detected in \ha\
emission with well-defined shell walls.  Diffuse X-ray emission is
detected toward DEM~L~152 (See Figure~\ref{fig:contour}b).
Additionally, diffuse X-ray emission is detected $\sim$6$'$ to
the northeast of DEM~L~152; this emission has been identified by
\citeauthor*{Paper2} as originating from a supernova remnant.

The diffuse X-ray emission correlates well with the \ha\ shell of
DEM~L~152.  The X-ray emission is limb-brightened, forming an X-ray
shell just interior to the \ha\ shell.  A breakout region on the
southern edge of the superbubble and an X-ray blister on the eastern
edge of the shell are detected in the PSPC image as well.  A more
thorough interpretation of the X-ray emission and the breakout regions
can be found in \citeauthor*{Paper2} and \citeauthor*{Paper4}.

\subsection{N51 (DEM~L~192, DEM~L~205)}
\label{sec:n51}
N51 is a nebular complex encompassing five OB associations, LH~51,
LH~54, LH~55, LH~60, and LH~63 (\citeauthor*{Paper1}).  Two ionized
gas shells are visible in the \ha\ image (see
Figure~\ref{fig:contour}c) and are cataloged as DEM~L~192 and
DEM~L~205.  DEM~L~192 is the larger shell with a size of 135~pc
$\times$ 120~pc and contains the OB associations LH~51 and LH~54.
DEM~L~205 is the smaller shell with a size of 65~pc $\times$ 50~pc.
The morphology of DEM~L~205 is that of a blister with the OB
association LH~63 at the base.  Diffuse X-ray emission is detected
toward both of these optical shells.

The diffuse X-ray emission toward DEM~L~192 is limb-brightened and
confined within the optical \ha\ shell.  Therefore, it is reasonable
to conclude that the X-ray emission is produced by hot gas interior to
the superbubble.  Additionally, an X-ray hotspot is detected within
the superbubble and coincides with the Wolf-Rayet star Br31, also
cataloged as Sk$-$67$^{\circ}$104.  Emission from this hotspot was
excluded in the thermal plasma model fit of DEM~L~192.

It is not clear whether the X-ray hotspot is a peak in the diffuse
emission or a stellar source.  Only 42$\pm$11 PSPC counts were
detected from the hotspot.  This count level is inadequate to
constrain the three parameters upon which a thermal plasma emission
model depends.  Further high resolution X-ray observations are needed
to explore the nature of this hotspot.

DEM~L~205 also shows limb-brightening in the PSPC images.  Unlike
DEM~L~192, no X-ray hotspots are detected within DEM~L~205.  On the
southwestern side of this shell, where the \ha\ surface brightness is
lowest, the X-ray emission appears more extended than the main shell.
X-ray emission of similar morphology is also detected in the Einstein
observations of N51 (\citeauthor*{Paper1}).  This may indicate a
breakout of the hot gas interior to DEM~L~205 into a lower density
region; however, the \ha\ image hints at the presence of very faint
outer \ha\ filaments confining this emission.

\subsection{N57 (DEM~L~229)}
\label{sec:n57}
N57 is a nebular complex encompassing the OB association LH~76.  Two
ionized gas shells are visible in the \ha\ image (See
Figure~\ref{fig:contour}d) and are cataloged as DEM~L~229 and
DEM~L~231.  DEM~L~229 is the larger shell; it is 200~pc $\times$
100~pc in size and contains LH~76.  The smaller shell, DEM~L~231,
30~pc $\times$ 30~pc in size, is a ring nebula around the Wolf-Rayet
star Br48 \citep{Chu80,CWG99}.  Of these two shells, diffuse X-ray
emission is only detected toward DEM~L~229.  Additionally, a bright
X-ray source is visible to the north of DEM~L~229, the X-ray source is
not associated with DEM~L~229 and has been identified as the SNR
candidate, SNR 0532-675 \citep{Mat85,Williams99}.

The diffuse X-ray emission toward DEM~L~229 is coincident with the
interior of the superbubble.  The emission also appears to be largely
confined by the optical \ha\ shell.  This indicates that the X-ray
emission originates from the hot gas in the superbubble interior.  The
X-ray emission appears brightest towards the southern part of
DEM~L~229.  Details of the X-ray morphology cannot be confidently
determined, however, as N57 is detected on the outer edge of the PSPC
(See Table~\ref{tbl:PSPCobs}) where the point-spread function becomes
quite poor.

\subsection{N103 (DEM~L~84)}
\label{sec:n103}
N103 is a nebular complex encompassing the star cluster NGC\,1850.  It
has two main components, N103A (DEM~L~85), the 20~pc $\times$ 15~pc
\ha\ knot on the east, and N103B\footnote{It should be noted here that
the label ``N103B'' is often erroneously used to refer to the
supernova remnant 0509$-$68.7.  We are using ``N103B'' to refer to the
\ha\ structure as originally identified by \citet{Henize56}.}
(DEM~L~84), a 120~pc $\times$ 120~pc superbubble on the west.
Supernova remnant 0509$-$68.7 is present just exterior to the eastern
edge of the superbubble \citep{Mat83,Williams99}, and the superbubble
is brightest in \ha\ on the eastern side closest to the SNR.  Diffuse
X-ray emission is detected toward both the superbubble and the
supernova remnant (See Figure~\ref{fig:contour}e).

The X-ray emission toward N103B appears to have several distinct
structures.  The most obvious is the very strong extended source that
is coincident with the supernova remnant 0509$-$68.7.  Another
prominent structure is a large limb-brightened X-ray ring with a
point X-ray source at the center.  \citet{Chu00} found that this
X-ray ring is centered on the nearby star cluster HS122 \citep{HS66},
which is projected at the southwest rim of the superbubble.  We
therefore cannot assume that the majority of the diffuse X-ray
emission detected toward N103B is caused by hot gas inside the
superbubble of N103B.  Further comparison of the X-ray and \ha\ images
presented in Figure~\ref{fig:contour}e reveal diffuse X-ray emission
in the region between the supernova remnant of N103B and the X-ray
ring and coincident with the superbubble.  Because of the PSPC's low
scattering levels, it is reasonable to assume that this X-ray emission
arises from hot gas within the superbubble.  We have used this region
as a sample of the X-ray emission from the superbubble of N103B.

\subsection{N105 (DEM~L~86)}
\label{sec:n105}
N105 is a nebular region encompassing the OB association LH~31 (also
cataloged as NGC\,1858) and the star cluster NGC\,1854.  Two bright
\ha\ knots are visible as well as \ha\ emission from the filaments of
the 90~pc $\times$ 60~pc superbubble shell.  The larger \ha\ knot is
coincident with LH~31; the smaller knot is coincident with NGC\,1854.
Diffuse X-ray emission is detected toward the larger \ha\ knot (See
Figure~\ref{fig:contour}f).

The strongest X-ray emission from N105 is coincident with LH~31,
indicating that the X-ray emission is likely emitted by hot gas
produced by the OB association.  The diffuse X-ray is not confined by
the \ha\ emission of N105; instead, the emission extends eastward from
N105 to a nearby patch of \ha\ emission, DEM~L~87.  The diffuse X-ray
emission appears to lose intensity approximately at the edge of
DEM~L~87.  This suggests that the hot, X-ray emitting gas created by
LH~31 is expanding eastward through a lower density medium to the
denser gas in DEM~L~87.  At the interface with DEM~L~87, the X-ray
emission drops off, further suggesting interaction between the X-ray
emitting gas of N105 and DEM~L~87.

\subsection{N144 (DEM~L~199)}
\label{sec:n144}
N144, encompassing the OB association LH~58, is a nebular region near
the western rim of the supergiant shell LMC 3.  A 120~pc $\times$
75~pc ionized gas shell is visible in the \ha\ image (See
Figure~\ref{fig:contour}g).  The morphology of N144 is a roughly
circular complex made up of many blister-like bubbles surrounding a
central shell structure.  The central shell has the brightest \ha\
emission of the nebula on its northern side.  Diffuse X-ray emission
is weakly detected toward several of the bubble regions, including the
central bubble.

The X-ray emission is generally coincident with N144, and the
strongest X-ray emission is toward the central shell seen in the \ha\
image.  This indicates that the X-ray emission is likely to originate 
from the hot gas in the superbubble interior.  There are two peaks
in the X-ray emission from the central shell.  These peaks are generally
coincident with the Wolf-Rayet stars Br34 and Br32,
also cataloged as Sk$-$68$^{\circ}$82 and Sk$-$68$^{\circ}$80,
indicating probable sources for the peaks in the diffuse
X-ray emission \citep*{Hail01}.  The X-ray emission 
extends beyond the \ha\ emitting shell of the central bubble to the 
northeast and southwest.  Each of these ``wings'' from central
concentration has its own weaker concentration of X-ray 
emission. This suggests that multiple-bubble structures in 
the N144 region have interior hot diffuse gas.  The southwest 
``wing'' of the X-ray emission shows further extension beyond the 
\ha\ shell of N144.  This may indicate a breakout of the 
hot gas interior to N144 into a lower density region.

\subsection{N154 (DEM~L~246)}
\label{sec:n154}
N154 is a nebular complex to the south of the 30 Doradus region.  The
superbubble encompasses the OB associations LH~81 and LH~87.  The
180~pc $\times$ 120~pc \ha\ shell of the superbubble is an angular,
almost-rhomboid shape, with the strongest \ha\ emission coming from
the northeast and southwest sides.  Diffuse X-ray emission is detected
toward N154 (see Figure~\ref{fig:contour}h).  A bright X-ray source to
the southwest of N154 has been identified as SNR 0534-699
\citep{Mat83,Williams99}.

The diffuse X-ray toward N154 is centrally bright, and the morphology
of the emission is that of an ellipsoidal running southwest to
northeast - similar to the structure and size of the \ha\ emission.
It is reasonable to therefore assume that the diffuse X-ray emission
arises from hot gas in the interior of the superbubble N154.  There is
significant X-ray emission trailing from N154 toward the northeast.
Due to the strong X-ray background emission and number of X-ray
sources near the 30 Doradus complex, it is difficult to determine if
this emission trail is a breakout region of hot X-ray emitting gas
from N154 or a blending of X-ray emission from N154 with emission from
neighboring regions.

\subsection{N158 (DEM~L~269)}
\label{sec:n158}
N158 is a complex nebular structure to the south of the 30~Doradus.
The complex encompasses the OB associations LH~101 and LH~104.  The
superbubble is located on the northern part of the N158 complex and
has a well-defined, 100~pc $\times$ 90~pc shell in the \ha.  Diffuse
X-ray emission is detected toward N158 (See
Figure~\ref{fig:contour}i).  Two strong X-ray sources are detected to
the north and west, respectively, of N158 as well.  The source north
of N158 appears to be coincident with the known SNR and pulsar
PSR~B0540$-$69.3 \citep{Mat83,Bica98}, and the position of the source
west of N158 is consistent with the Einstein X-ray source
0538.5$-$6925, a foreground Galactic star \citep{CSM97}.

The diffuse X-ray emission toward N158 appears to be coincident with
the superbubble.  Although the X-ray emission is not confined by the
observed \ha\ shell, the morphology of the X-ray emission suggests
that the emission is associated with N158, but that several breakout
regions have formed along the shell where hot, X-ray emitting gas is
escaping the interior of the superbubble.  It must be cautioned,
however, that the X-ray morphology of this region is exceedingly
complex.

\subsection{N160 (DEM~L~284)}
\label{sec:n160}
The nebular complex N160 is located on the southern side of the 30
Doradus region.  The superbubble in N160 dominates the morphology of
the complex.  The superbubble is 180~pc $\times$ 150~pc in size, and
encompasses the OB association LH~103.  The \ha\ morphology is roughly
circular, with a possible ``blowout'' region apparent on the northeast
edge \citep{Points99}.  The \ha\ emission is strongest on the southern
edge of the superbubble, closest to LH~103.  Diffuse X-ray emission is
detected toward the superbubble (See Figure~\ref{fig:contour}j).
Additionally, there is a pair of strong X-ray sources south of N160.
These sources have been identified: the brighter source is the X-ray
binary LMC~X-1 and the dimmer source is SNR~0540-697
\citep{Chu97,Williams00}.

The diffuse X-ray emission detected toward the superbubble in N160 is
concentrated near LH~103.  Little significant emission is detected
from the remainder of the the superbubble.  This suggests that the
stars and/or supernovae in LH~103 are producing hot X-ray emitting
gas.  The gas interior to the superbubble may be too hot to be
detected in the PSPC energy bandpass, or the majority of the hot gas
may simply have already escaped.

\subsection{N206 (DEM~L~221)}
\label{sec:n206}
N206 (also cataloged as DEM~L~221) is a nebular complex encompassing
the OB associations LH~66 and LH~69.  N206 contains both a superbubble
and a smaller supernova remnant, SNR 532-710 \citep{Mat83,Williams99}.
The 30~pc $\times$ 30~pc remnant is located on the eastern side of the
nebular complex, and has a faint circular \ha\ shell.  The superbubble
has a larger circular shell, 110~pc $\times$ 110~pc, with the
brightest \ha\ emission coming from the eastern and southern sides of
the bubble.  Diffuse X-ray emission is detected toward both the
remnant and the superbubble (See Figure~\ref{fig:contour}k).  The
diffuse X-ray emission detected toward the supernova remnant has been
previously explored by \citet{Williams99}.

The diffuse X-ray emission toward the superbubble of N206 appears to
have a several enhancements, possibly in a limb-brightened ring.  The
emission is coincident with a region of the superbubble that is not
bright in \ha, but the diffuse X-ray emission appears confined by
faint \ha\ structures on the western side of the superbubble.
Therefore, it is reasonable to conclude that the X-ray emission is
produced by hot gas interior to the superbubble.

One of the enhancements of the diffuse X-ray emission is
coincident with the Wolf-Rayet star Br44.  The X-ray emission from
this enhancement in the superbubble may therefore have a stellar
source rather than a diffuse one.  The emission from this enhancement
has therefore been excluded from the thermal plasma model fit of the
superbubble in N206.

\subsection{30~Dor~C (DEM~L~263)}
\label{sec:30dorc}
30~Dor~C is a superbubble located in the southwestern region of the 30
Doradus complex.  The superbubble encompasses the OB association
LH~90.  The \ha\ emission shows a strong shell structure, 100~pc
$\times$ 90~pc in size.  Diffuse X-ray emission is detected toward the
superbubble (See Figure~\ref{fig:contour}l).

The X-ray emission detected toward 30~Dor~C is limb-brightened and
appears confined within the \ha\ shell.  It is therefore reasonable to
conclude that the X-ray emission arises from within the superbubble.
The limb-brightened shell is well-defined all around the superbubble
except on the southwest, where the X-ray emission is near the
background levels.

The absorption column density towards 30~Dor~C is known to change
dramatically across the face of the superbubble \citep{Ost97}.  We
have therefore divided 30~Dor~C into two parts (east and west) to
account for the change in absorption column density in our spectral
fits.  Unfortunately, the LMC $N_{\rm HI}$ maps from \citet{Rohlfs84}
do not have the spatial resolution to detect this change, so we must
still use a single value in our ``fixed $N_{\rm H}$'' fits of
30~Dor~C.

\section{Discussion}

\subsection{Distribution of X-ray Luminosities}
Based on the spectral fits performed for each superbubble, we have
computed X-ray luminosities.  We have used these luminosities to
investigate the superbubble luminosity function.  In
Figure~\ref{fig:LumCurve}, we present the X-ray luminosity function
for the superbubbles in our dataset.  We have plotted the luminosities
as determined by both the ``best-fit $N_{\rm H}$'' model fits and the
``fixed $N_{\rm H}$'' model fits.  The ``best-fit $N_{\rm H}$'' luminosity
function illustrates that most of the superbubbles in our sample have
an X-ray luminosity around 10$^{35-36}$ ergs sec$^{-1}$.  A single
high-end outlier is also shown at over 10$^{37}$ ergs sec$^{-1}$; this
outlier is N11--Shell 1.  The ``fixed $N_{\rm H}$'' luminosity distribution
favors higher X-ray luminosities than the ``best-fit $N_{\rm H}$''
luminosity distribution as well as a broader range in luminosities (excepting
the outlier on the ``best-fit $N_{\rm H}$'' luminosity function).

\subsection{Pressure Driven Models}
To compare the X-ray luminosities of the superbubbles derived from
observation with the pressure-driven bubble models of \citet{Weaver77}
for the LMC superbubbles, we have followed the procedure described in
\citeauthor*{Paper1} and corrected in \citeauthor*{Paper3}.  Assuming
that the shell thickness is small compared to the radius of the
bubble, an electron temperature in the shell of $T_{e} \simeq$ 10$^4$
K, and a mean atomic mass of the ambient medium $\mu_{a}$ =
(14/11)$m_{\rm H}$, we can use equations (7)--(10) from
\citeauthor*{Paper3} to derive the X-ray luminosity:

\begin{equation}
L_{\rm X} \simeq (6.7 \times 10^{29} {\rm erg} \cdot {\rm sec}^{-1})
\cdot \xi \cdot I \cdot EM^{5/7} \cdot R_{\rm pc}^{12/7} 
\cdot v_{\rm exp}^{1/7}
\end{equation}

\noindent where $\xi$ is the metallicity, assumed to be 0.3, $I$ is a
dimensionless function of the temperatures interior to the
superbubble, which has a value $\sim$2, $EM$ is the emission measure
of the 10$^4$ K shell gas in cm$^{-6}$ pc, $R_{\rm pc}$ is the radius
of the superbubble in parsecs, and $v_{\rm exp}$ is the superbubble
expansion velocity in km sec$^{-1}$.  The dimensions and expansion
velocities of the superbubbles are given in Table~\ref{tbl:Prop}.
Unfortunately, we do not have expansion velocities for all of the
superbubbles.  The emission measure was determined from the
continuum-subtracted \ha\ image derived from the PDS scans of the
Curtis-Schmidt plates of \citet{KH86}.  The emission measures and
theoretical X-ray luminosities are presented in Table~\ref{tbl:EM}.
Of course, the emission measure can vary greatly around \ha\ shell, so
we have taken an rough mean for each superbubble.  The predicted X-ray
luminosities range from 10$^{34.3}$--10$^{35.1}$ erg sec$^{-1}$.
These luminosities range from $\sim$3 to $\sim$50 times lower than the
X-ray luminosities determined from the PSPC data with the ``fixed
$N_{\rm H}$'' fits.  This suggests that the majority of the X-rays are
produced by different mechanisms than those described in the
pressure-driven bubble model, confirming that the superbubbles in our
sample are X-ray bright.

\subsection{X-Ray Luminosity Correlations}
We have also compared the superbubble volume, \ha\ luminosity,
expansion velocity, and bright star count with the X-ray luminosity,
for both ``best-fit $N_{\rm H}$'' and ``fixed $N_{\rm H}$'' model fits
(See Figures~\ref{fig:LumVol}, \ref{fig:LumHA}, \ref{fig:LumExp}, \&
\ref{fig:LumOBStar}).  The superbubble volumes were determined from
the sizes of the \ha\ shell (See Table~\ref{tbl:Prop}), assuming an
ellipsoidal shape.  The expansion velocities and \ha\ luminosities are
also given in Table~\ref{tbl:Prop}. The bright star counts are based
on the OB association star counts in \citet{LH70}; again,
Table~\ref{tbl:Prop} lists the OB associations encompassed by each
superbubble.  Although the scatter level of these plots is obviously
high, we have attempted to fit each plot with a linear trend line to
test for correlations between X-ray luminosity and other superbubble
properties.  The correlation coefficients of the trends are detailed
in Table~\ref{tbl:Corr}.  Positive correlations are found between
X-ray luminosity and each of the other properties.  The correlations
are moderate for the X-ray luminosities as determined by the
``best-fit $N_{\rm H}$'' model fits and generally stronger for the
X-ray luminosities as determined by the ``fixed $N_{\rm H}$'' model
fits.  The strongest correlation is between ``fixed $N_{\rm H}$''
X-ray luminosity and bright star count.  It must be considered,
however, that the correlation between X-ray luminosity and superbubble
volume may be due to a surface brightness selection effect.

The correlations demonstrate that the X-ray luminosity of a
superbubble is affected by the richness and age of the OB associations
within its shell walls.  The bright star count of a superbubble will
obviously be directly related to the richness of its OB associations.
Also, OB association richness will provide stellar winds to power the
expansion, and thereby increase the expansion velocity, of the
superbubble.  The \ha\ luminosity of a superbubble will be positively
affected by the richness of OB association, as more stars provide more
ionizing flux, and negatively affected by age, as the
powerfully-ionizing, early-type stars exhaust themselves.  The X-ray
luminosity--OB association richness relationship has already been
demonstrated; however, the X-ray luminosity can increase with the age
of a superbubble.  As demonstrated by \citeauthor*{Paper1} and
\citet{Wang91}, the X-ray luminosity of a superbubble can be enhanced
by SNRs.  Thus, a superbubble that has already had several bright
stars go supernova can be brighter in X-rays than a superbubble with
much younger OB associations.  We would therefore expect the
correlation between X-ray luminosity and \ha\ luminosity to be weaker
than the correlation between X-ray luminosity and bright star count,
which it is for the ``fixed $N_{\rm H}$'' model fits.

\subsection{Stellar Sources and Breakout Regions}
We have described the X-ray morphology for each superbubble and
compared those morphologies to the \ha\ morphologies and known stellar
sources of each superbubble.  We have found that in a significant
fraction of the superbubbles, peaks in the X-ray emission are
coincident with known stellar sources, such as Wolf-Rayet stars.
High-resolution observations are need to determine whether the X-ray
peaks are caused by stellar emission or stellar wind interactions with
the superbubble interior gas.  In addition, nearly half of the
superbubbles show some evidence of breakout regions in their X-ray
morphologies, where hot gas appears to be leaking from the superbubble
interior into the surrounding regions.  Again, further studies of the
diffuse X-ray gas will be needed to confirm whether these regions are
true breakout regions.

\section{Summary}
\label{sec:Sum}

We have presented {\it ROSAT} observation of thirteen LMC
superbubbles.  Eleven of these observations had not been reported
previously.  In each of these superbubbles, diffuse X-ray emission
brighter than is theoretically expected for a wind-blown bubble was
detected.  Based on the previous findings in \citeauthor*{Paper1} and
\citet{Wang91}, it is reasonable to conclude that the X-ray emission
from the superbubbles has been enhanced by interactions between the
superbubble shell and interior SNRs.  We have also found significant
positive correlations between the X-ray luminosity of a superbubble
and its \ha\ luminosity, expansion velocity and OB star count.  
Further, we have
found that a large fraction of the superbubbles in the sample show
evidence of breakout regions.  In \citeauthor*{Paper4} it was
demonstrated that breakout regions can significantly affect the
evolution of a superbubble, draining energy and pressure that would
otherwise go into expansion.  We also suggest that because these
breakout regions appear so frequently, the superbubbles may be a
significant source of hot gas for the interstellar medium.

\acknowledgments{We would like to thank Robert Gruendl for his useful
communications in preparing this paper.  This research has made use of
data obtained through the High Energy Astrophysics Science Archive
Research Center Online Service, provided by the NASA/Goddard Space
Flight Center.  This research was made possible by ADP grants NAG
5-7003 and NAG 5-8003.}

\clearpage

\figcaption[]{For each pair, the left image show \ha\ emission from the
superbubble(s) overlaid with X-ray contours.  The right image, shows
the X-ray emission overlaid with the same contours to ensure the
clarity of the contour levels.  The contours are at levels of 50\%,
70\%, 85\%, and 95\% of the peak level within the superbubble.  For
bright X-ray objects in the field not actually part of the superbubble
(such as SNRs), we have plotted additional contours at 2, 4, 8, and 16
times the superbubble peak level.  These contours are plotted as
dashed lines.  Locations of the superbubbles are indicated in the \ha\
images.  (a) N11--Shell 1; (b) N44; (c) N51; (d) N57; (e) N103; (f)
N105; (g) N144; (h) N154; (i) N158; (j) N160; (k) N206; and (l)
30~Dor~C. \label{fig:contour}}

\figcaption[]{Each plot shows the superbubble's X-ray
spectra and the ``best-fit $N_{\rm H}$'' spectral fit.  (a) N11--Shell
1; (b) N44; (c) N51--DEM~L~192; (d) N51--DEM~L~205.  (e) N57; (f)
N103; (g) N105; (h) N144; (i) N154; (j) N158; (k) N160; (l) N206.  (m)
30~Dor~C--east side; (n) 30~Dor~C--west side; \label{fig:freespec}}

\figcaption[]{Each plot shows the superbubble's X-ray
spectra and the ``fixed $N_{\rm H}$'' spectral fit.  (a) N11--Shell 1;
(b) N44; (c) N51--DEM~L~192; (d) N51--DEM~L~205.  (e) N57; (f) N103;
(g) N105; (h) N144; (i) N154; (j) N158; (k) N160; (l) N206.  (m)
30~Dor~C--east side; (n) 30~Dor~C--west side; \label{fig:fixspec}}

\figcaption[]{Superbubble X-ray luminosity distribution.
X-ray luminosity is plotted on a logarithmic scale.  The 0.2 bin size
was used. \label{fig:LumCurve}}

\figcaption[]{Superbubble X-ray luminosity vs. volume for
the ``best-fit $N_{\rm H}$'' X-ray spectral fits (upper) and ``fixed
$N_{\rm H}$'' X-ray spectral fits (lower).  Luminosity is plotted on a
logarithmic scale.  The volumes are based on \ha\ morphology.  A
linear trend line is plotted for each graph.  \label{fig:LumVol}}

\figcaption[]{Superbubble X-ray luminosity vs. \ha\
luminosity for the ``best-fit $N_{\rm H}$'' X-ray spectral fits
(upper) and ``fixed $N_{\rm H}$'' X-ray spectral fits (lower).  Both
luminosities are plotted on a logarithmic scale.  A linear trend line
is plotted for each graph.  \label{fig:LumHA}}

\figcaption[]{Superbubble X-ray luminosity vs. \ha\
expansion velocity for the ``best-fit $N_{\rm H}$'' X-ray spectral
fits (upper) and ``fixed $N_{\rm H}$'' X-ray spectral fits (lower).
Luminosity is plotted on a logarithmic scale.  A linear trend line is
plotted for each graph.  \label{fig:LumExp}}

\figcaption[]{Superbubble X-ray luminosity vs. bright
star count for the ``best-fit $N_{\rm H}$'' X-ray spectral fits
(upper) and ``fixed $N_{\rm H}$'' X-ray spectral fits (lower).
Luminosity is plotted on a logarithmic scale.  Bright star counts are
based on star counts in local OB associations
\citep{LH70}. \label{fig:LumOBStar}}

\clearpage
\pagestyle{empty}

\begin{deluxetable}{l l l c c c c c}
\tabletypesize{\footnotesize}
\tablecolumns{8}
\tablecaption{LMC Superbubble Properties \label{tbl:Prop}}
\tablehead{\colhead{} & \colhead{} & \colhead{} & \colhead{} &
	\colhead{} & \colhead{} & \colhead{} & \colhead{Expansion} \\
\colhead{Object\tablenotemark{a}} & \colhead{DEM No.\tablenotemark{b}} 
	& \colhead{Local OB\tablenotemark{c}}
	& \colhead{$\alpha_{2000}$} & \colhead{$\delta_{2000}$}
	& \colhead{Size\tablenotemark{d}} 
	& \colhead{log $L_{\rm H\alpha}$\tablenotemark{d}}
	& \colhead{Velocity} \\
  &  & \colhead{Associations} & \colhead{(h:m:s)} 
  	& \colhead{($^\circ$:':'')} & \colhead{(pc $\times$ pc)} 
	& \colhead{(ergs sec$^{-1}$)} & \colhead{(km sec$^{-1}$)}}
\startdata
N11--Shell 1&  DEM~L~ 34 & LH~9           & 04:56:51 & $-$66:24:24 & 150 $\times$ 100 & 36.90 & 45\tablenotemark{e} \\ 
N44         &  DEM~L~152 & LH~47          & 05:22:08 & $-$67:56:12 & 100 $\times$  75 & 37.00 & 40\tablenotemark{f} \\
N51         &  DEM~L~192 & LH~51, LH~54   & 05:26:14 & $-$67:30:18 & 135 $\times$ 120 & 37.04 & 20--50\tablenotemark{g} \\
            &  DEM~L~205 & LH~63          & 05:28:06 & $-$67:28:36 &  65 $\times$  50 & 36.26 & 45--70\tablenotemark{g} \\
N57         &  DEM~L~229 & LH~76          & 05:32:24 & $-$67:41:18 & 135 $\times$ 105 & 36.91 & $\sim$45\tablenotemark{g} \\
N103B       &  DEM~L~ 84 & NGC\,1850      & 05:08:54 & $-$68:45:00 & 120 $\times$ 120 & 36.53 & 20\tablenotemark{h} \\
N105        &  DEM~L~ 86 & LH~31          & 05:09:57 & $-$68:53:31 &  90 $\times$  60 & 36.82 & -- \\
N144        &  DEM~L~199 & LH~58          & 05:26:33 & $-$68:51:48 & 120 $\times$  75 & 36.98 & 20--30\tablenotemark{g} \\
N154        &  DEM~L~246 & LH~81, LH~87   & 05:35:57 & $-$69:38:54 & 180 $\times$ 120 & 37.10 & -- \\
N158        &  DEM~L~269 & LH~101, LH~104 & 05:39:33 & $-$69:25:48 & 120 $\times$ 100 & 36.80 & $\sim$45\tablenotemark{g} \\
N160        &  DEM~L~284 & LH~103         & 05:40:12 & $-$69:37:06 & 180 $\times$ 150 & 37.22 & $\lesssim$20\tablenotemark{g} \\
N206        &  DEM~L~221 & LH~66, LH~69   & 05:29:36 & $-$71:00:00 &  90 $\times$  90 & 36.95 & $\sim$30\tablenotemark{g} \\
30~Dor~C    &  DEM~L~263 & LH~90          & 05:38:42 & $-$69:06:03 & 100 $\times$  90 & 36.94 & $\sim$45\tablenotemark{g} \\
\enddata
\tablenotetext{a}{Nomenclature of \citet{Henize56}, except 30~Dor~C}
\tablenotetext{b}{Nomenclature of \citet{DEM76}}
\tablenotetext{c}{Nomenclature of \citet{LH70}, except NGC\,1850}
\tablenotetext{d}{Determined from PDS scans of \citet{KH86}, includes only \ha\ emission from the superbubble shell} 
\tablenotetext{e}{\citet{Rosado96}}
\tablenotetext{f}{\citet{ML91}}
\tablenotetext{g}{\citet{ChuInPrep}}
\tablenotetext{h}{\citet{Georgelin83}}
\end{deluxetable}

\clearpage
\pagestyle{empty}

\begin{deluxetable}{l l l c c r r}
\tabletypesize{\footnotesize}
\tablecolumns{7}
\tablecaption{Archival {\it ROSAT} Observations \label{tbl:PSPCobs}}
\tablehead{\colhead{Field(s)} & \colhead{{\it ROSAT} ID} & \colhead{PI} 
	& \colhead{$\alpha_{2000}$} & \colhead{$\delta_{2000}$}
	& \colhead{Exposure Time} & \colhead{Off-axis Angle} \\
  &  &  & \colhead{(h:m:s)} & \colhead{($^\circ$:$'$:$''$)} 
  	& \colhead{(sec)} & \colhead{(arcmin)}}
\startdata
N11          & rp900320a01 & Chu        & 04:56:33.6 & $-$66:28:48 & 13731 & 4.7 \\
             & rp900320n00 & Chu        & 04:56:33.6 & $-$66:28:48 & 17589 & 4.7 \\
N44          & rp400154n00 & Pakull     & 05:22:26.4 & $-$67:58:12 &  6522 & 2.2 \\
             & rp500093n00 & Chu        & 05:22:02.4 & $-$67:55:12 &  8720 & 2.2 \\
N51, N57     & rp500054a00 & Fink       & 05:25:52.8 & $-$67:30:00 &  3420 & 5.3, 39.7 \\
             & rp500054a01 & Fink       & 05:25:52.8 & $-$67:30:00 &  4452 & 5.3, 39.7 \\
N103, N105   & rp500037n00 & Aschenbach & 05:08:60.0 & $-$68:43:48 &  6826 & 1.3, 11.0 \\
N144         & rp500138a01 & MacLow     & 05:26:36.0 & $-$68:50:24 & 14531 & 1.4 \\
             & rp500138a02 & MacLow     & 05:26:36.0 & $-$68:50:24 & 14581 & 1.4 \\
             & rp500138n00 & MacLow     & 05:26:36.0 & $-$68:50:24 &  2478 & 1.4 \\
N154, 30~Dor~C & rp180251n00 & Aschenbach & 05:35:28.8 & $-$69:16:12 & 20153 & 22.8, 13.4 \\
             & rp500100a00 & Gorenstein & 05:35:28.8 & $-$69:16:12 & 16957 & 22.8, 13.4 \\
             & rp500100a01 & Gorenstein & 05:35:28.8 & $-$69:16:12 &  9657 & 22.8, 13.4 \\
             & rp500140a02 & Gorenstein & 05:35:28.8 & $-$69:16:12 & 10758 & 22.8, 13.4 \\
             & rp500140n00 & Gorenstein & 05:35:28.8 & $-$69:16:12 &  2642 & 22.8, 13.4 \\
             & rp500303n00 & Hasinger   & 05:35:28.8 & $-$69:16:12 &  9416 & 22.8, 13.4 \\
N158         & rp400052n00 & Oegelman   & 05:40:12.0 & $-$69:19:48 &  8823 & 10.9 \\
             & rp400133n00 & Oegelman   & 05:40:12.0 & $-$69:19:48 &  1803 & 10.9 \\
N160         & rp400079n00 & Lewin      & 05:39:38.4 & $-$69:44:24 &  7429 & 7.9 \\
N206         & rp300172a01 & Oegelman   & 05:32:28.8 & $-$70:21:36 &  2993 & 41.0 \\
             & rp300172a02 & Oegelman   & 05:32:28.8 & $-$70:21:36 &  3880 & 41.0 \\
             & rp300172n00 & Oegelman   & 05:32:28.8 & $-$70:21:36 &  6272 & 41.0 \\
             & rh600781a01 & Chu        & 05:30:45.6 & $-$71:02:24 & 24041 & 6.1 \\
             & rh600781n00 & Chu        & 05:30:45.6 & $-$71:02:24 & 25238 & 6.1 \\
\enddata
\end{deluxetable}

\clearpage
\pagestyle{empty}

\begin{deluxetable}{l r r r c c c c}
\tabletypesize{\footnotesize}
\tablecolumns{8}
\tablecaption{``Best-Fit $N_{\rm H}$'' LMC Superbubble X-ray Spectral Fits \label{tbl:FreeSpec}}
\tablehead{\colhead{} & \colhead{} & \colhead{Background} & \colhead{Scaled}
	& \colhead{} & \colhead{} & \colhead{} & \colhead{} \\
\colhead{Object} & \colhead{Exposure} & \colhead{Subtracted} 
	& \colhead{Background}
        & \colhead{log $N_{\rm H}$} & \colhead{$kT$}
	& \colhead{log $L_{\rm X}$\tablenotemark{a}} 
	& \colhead{N$_{\rm e} \sqrt{f}$} \\
\colhead{} & \colhead{(sec)} & \colhead{Source Counts} & \colhead{Counts}
	& \colhead{(cm$^{-2}$)} & \colhead{(keV)}  
	& \colhead{(erg s$^{-1}$)} & \colhead{(10$^{-2}$ cm$^{-3}$)}}
\startdata
N11--Shell 1 & 31320 &  1775 & 1667 & 21.9 & 0.13 & 37.17 & 31. \\ 
N44          & 15242 &  2894 & 1260 & 20.6 & 0.71 & 35.73 & 3.8 \\
N51 \\
~~~DEM~L~192 &  7872 &   559 &  865 & 20.4 & 0.28 & 35.18 & 1.5 \\
~~~DEM~L~205 &  7872 &   220 &  226 & 20.5 & 0.25 & 34.86 & 3.7 \\
N57          &  7872 &   398 &  489 & 20.6 & 0.33 & 35.32 & 1.8 \\
N103B        &  6826 &    99 &  120 & 21.3 & 0.72 & 34.85 & 0.8 \\ 
N105         &  6826 &   158 &  183 & 21.6 & 0.31 & 35.40 & 4.6 \\
N144         & 31590 &   667 & 1399 & 20.5 & 0.32 & 34.74 & 1.3 \\
N154         & 81210 &  6269 & 9618 & 20.7 & 0.29 & 35.52 & 1.8 \\
N158         & 10626 &  2543 & 1236 & 20.7 & 0.60 & 35.85 & 3.0 \\ 
N160         &  7429 &   964 & 1517 & 20.9 & 3.40 & 35.80 & 1.9 \\
N206         & 13145 &   330 &  490 & 20.7 & 0.31 & 35.09 & 2.2 \\ 
30~Dor~C \\
~~~East      & 81210 &  5378 & 4388 & 21.1 & 1.19 & 35.49 & 4.0 \\
~~~West      & 81210 &  1887 & 3225 & 22.0 & 1.22 & 35.84 & 7.4 \\
\enddata
\tablenotetext{a}{In the energy band 0.5--2.4 keV.}
\end{deluxetable}

\clearpage
\pagestyle{empty}

\begin{deluxetable}{l c c c c c c}
\tabletypesize{\normalsize}
\tablecolumns{7}
\tablecaption{``Fixed $N_{\rm H}$'' LMC Superbubble X-ray Spectral Fits \label{tbl:FixSpec}}
\tablehead{\colhead{Object} & \colhead{log $(N_{\rm HI})_{\rm Galactic}$\tablenotemark{a}} 
	& \colhead{log $(N_{\rm HI})_{\rm LMC}$\tablenotemark{b}}  
        & \colhead{log $N_{\rm H}$} 
        & \colhead{$kT$} & \colhead{log $L_{\rm X}$\tablenotemark{c}} 
	& \colhead{N$_{\rm e} \sqrt{f}$}\\
\colhead{} & \colhead{(cm$^{-2}$)} & \colhead{(cm$^{-2}$)} 
        & \colhead{(cm$^{-2}$)} & \colhead{(keV)}  
	& \colhead{(erg s$^{-1}$)} & \colhead{(10$^{-2}$ cm$^{-3}$)}}
\startdata
N11--Shell 1 & 20.63 & 21.45 & 21.6 & 0.24 & 35.96 & 4.2 \\ 
N44          & 20.79 & 21.45 & 21.6 & 0.35 & 36.36 & 9.3 \\
N51 \\
~~~DEM~L~192 & 20.78 & 21.07 & 21.4 & 0.18 & 35.82 & 4.1 \\
~~~DEM~L~205 & 20.77 & 20.89 & 21.3 & 0.18 & 35.37 & 8.2 \\
N57          & 20.78 & 21.23 & 21.5 & 0.23 & 35.91 & 4.3 \\
N103B        & 20.79 & 21.30 & 21.5 & 0.65 & 35.02 & 1.0 \\ 
N105         & 20.81 & 21.30 & 21.5 & 0.34 & 35.28 & 3.5 \\
N144         & 20.79 & 20.87 & 21.3 & 0.25 & 35.12 & 2.3 \\
N154         & 20.82 & 21.58 & 21.7 & 0.15 & 36.88 & 14. \\
N158         & 20.83 & 21.52 & 21.7 & 0.28 & 36.66 & 10. \\ 
N160         & 20.84 & 21.53 & 21.7 & 0.93 & 36.07 & 2.4 \\
N206         & 20.84 & 21.27 & 21.5 & 0.20 & 35.80 & 6.1 \\ 
30~Dor~C \\
~~~East      & 20.81 & 21.63 & 21.7 & 0.79 & 35.90 & 5.4 \\
~~~West      & 20.81 & 21.63 & 21.7 & --\tablenotemark{d} & 35.54 & 5.1 \\
\enddata
\tablenotetext{a}{\citet{DL90}}
\tablenotetext{b}{\citet{Rohlfs84}}
\tablenotetext{c}{In the energy band 0.5--2.4 keV.}
\tablenotetext{d}{Unable to Find Solution} 
\end{deluxetable}

\clearpage
\pagestyle{empty}

\begin{deluxetable}{l c c c c c c}
\tabletypesize{\footnotesize}
\tablecolumns{7}
\tablecaption{Superbubble Emission Measures and Pressure-Driven Model 
X-Ray Luminosities \label{tbl:EM}}
\tablehead{\colhead{} & \colhead{Mean Emission} & \colhead{}
	& \colhead{Expansion} & \colhead{} & \colhead{} & \colhead{} \\
\colhead{Object} & \colhead{Measure}
	& \colhead{Size} & \colhead{Velocity} 
	& \colhead{log $(L_{\rm X})_{\rm Model}$} 
	& \colhead{log $(L_{\rm X})_{\rm Best-Fit}$}
	& \colhead{log $(L_{\rm X})_{\rm Fixed}$} \\
\colhead{} & \colhead{(cm$^{-6}$ pc)} & \colhead{(pc $\times$ pc)}
	& \colhead{(km sec$^{-1}$)} & \colhead{(erg sec$^{-1}$)}
	& \colhead{(erg sec$^{-1}$)} & \colhead{(erg sec$^{-1}$)}}
\startdata
N11--Shell 1 & $\sim$750  & 150 $\times$ 100 & 45       & 34.8 & 37.17 & 35.96 \\ 
N44          & $\sim$3000 & 100 $\times$  75 & 40       & 35.1 & 35.73 & 36.36 \\
N51 \\
~~~DEM~L~192 & $\sim$1750 & 135 $\times$ 120 & 20--50   & 34.8 & 35.18 & 35.82 \\
~~~DEM~L~205 & $\sim$1200 &  65 $\times$  50 & 45--70   & 35.1 & 34.86 & 35.37 \\
N57          & $\sim$1500 & 135 $\times$ 105 & $\sim$45 & 35.0 & 35.32 & 35.91 \\
N103B        & $\sim$1000 & 120 $\times$ 120 & 20       & 34.2 & 34.85 & 35.02 \\
N144         & $\sim$1500 & 120 $\times$  75 & 20--30   & 34.5 & 34.74 & 35.12 \\
N158         & $\sim$2000 & 120 $\times$ 100 & $\sim$45 & 35.0 & 35.85 & 36.66 \\ 
N160         & $\sim$1250 & 180 $\times$ 150 & $\lesssim$20 & 34.3 & 35.80 & 36.07 \\
N206         & $\sim$1500 &  90 $\times$  90 & $\sim$30 & 34.6 & 35.09 & 35.80 \\ 
30~Dor~C     & $\sim$2000 & 100 $\times$  90 & $\sim$45 & 35.0 & 36.00 & 36.06 \\
\enddata
\end{deluxetable}

\clearpage
\pagestyle{empty}

\begin{deluxetable}{l c c}
\tabletypesize{\normalsize}
\tablecolumns{3}
\tablecaption{Correlation Coefficients \label{tbl:Corr}}
\tablehead{\colhead{Plot} & \multicolumn{2}{c}{Correlation Coefficients} \\
\colhead{$L_{\rm X}$ vs.} & \colhead{Best-Fit $N_{\rm H}$} & \colhead{Fixed $N_{\rm H}$}} 
\startdata
Superbubble Volume & 0.30 & 0.40 \\
\ha\ Luminosity    & 0.33 & 0.51 \\
Expansion Velocity & 0.26 & 0.31 \\
Bright Star Count  & 0.35 & 0.83 \\
\enddata
\end{deluxetable}


\clearpage

\begin{thebibliography}{}

\bibitem[Arabadjis \& Bregman (1999)]{AB99} Arabadjis, J.S., \& 
Bregman, J.N. 1999, ApJ, 510, 806

\bibitem[Bica et al. (1998)]{Bica98} Bica, E.L.D., Schmitt, H.R., 
Dutra, C.M., Oliveira, H.L. 1998, AJ, 117, 238

\bibitem[Breysacher (1981)]{Brey81} Breysacher, J. 1981, A\&AS, 43, 203

\bibitem[Chu \& Lasker (1980)]{Chu80} Chu, Y.-H., \& Lasker, B.M. 1980,
PASP, 92, 730

\bibitem[Chu \& Mac Low (1990)Paper I]{Paper1} Chu, Y.-H., \& Mac Low, M.-M.
1990, ApJ, 365, 510 (Paper I)

\bibitem[Chu et al. (1993)Paper II]{Paper2} Chu, Y.-H., Mac Low, M.-M., 
Garcia-Segura, G., Wakker, B., Kennicutt, R.C. 1993, ApJ, 414, 213 (Paper II)

\bibitem[Chu et al. (1995)Paper III]{Paper3} Chu, Y.-H., Chang, H.-W., 
Su, Y.-L., \& Mac Low, M.-M. 1995, ApJ, 450, 157 (Paper III)

\bibitem[Chu et al. (1997)]{Chu97} Chu, Y.-H., Kennicutt, R.C., 
Snowden, S.L., Smith, R.C., Williams, R.M., \& Bomans, D.J. 1997
PASP, 109, 554

\bibitem[Chu, Weis, \& Garnet (1999)]{CWG99} Chu, Y.-H., Weis, K., \&
Garnett, D.R. 1999, AJ, 117, 1433

\bibitem[Chu et al. (2000)]{Chu00} Chu, Y.-H., Kim, S., Points, S.D.,
Petre, R., \& Snowden, S.L. 2000, AJ, 119, 2242

\bibitem[Chu et al. (2001)]{ChuInPrep} Chu, Y.-H. et al. 2001, in preparation

\bibitem[Cowley et al. (1997)]{CSM97} Cowley, A.P., Schmidtke, P.C.,
McGrath, T.K., Ponder, A.L., \& Fertig, M.R. 1997, PASP, 109, 21

\bibitem[Davies, Elliott, \& Meaburn (1976)]{DEM76} Davies, R.D., 
Elliott, K.H., \& Meaburn J. 1976, MmRAS, 81, 89

\bibitem[Dickey \& Lockman (1990)]{DL90} Dickey, J.M., \& Lockman, F.J.
1990, ARAA, 28, 215

\bibitem[Feast (1999)]{Feast99} Feast, M. 1999, PASP, 111, 775

\bibitem[Georgelin et al. (1983)]{Georgelin83} Georgelin, Y.M., 
Georgelin, Y.P., Laval, A., Monnet, G., Rosado, M. 1983, A\&AS, 54, 459

\bibitem[Hail, Dunne, \& Chu (2001)]{Hail01} Hail, T.C., Dunne, B.C, \& 
Chu, Y.-H. 2001, in preparation

\bibitem[Henize (1956)]{Henize56} Henize, K.G. 1956, ApJS, 2, 315

\bibitem[Hodge \& Sexton (1966)]{HS66} Hodge, P.W., \& Sexton, J.A. 1966,
AJ, 71, 363

\bibitem[Kennicutt \& Hodge (1986)]{KH86} Kennicutt, R.C., \& Hodge, P.W.
1986, ApJ, 306, 130

\bibitem[Lucke \& Hodge (1970)]{LH70} Lucke, P.B., \& Hodge P.W. 1970,
AJ, 75, 171

\bibitem[Mac Low et al. (1998)Paper V]{Paper5} Mac Low, M.-M., Chang, T.H.,
Chu, Y.-H., Points, S.D., Smith, R.C., Wakker, B.P. 1998, AJ, 493, 260
(Paper V)

\bibitem[Mathewson et al. (1983)]{Mat83} Mathewson, D.S., Ford, V.L.,
Dopita, M.A., Tuohy, I.R., \& Long, K.S. 1983, ApJS, 51, 345

\bibitem[Mathewson et al. (1985)]{Mat85} Mathewson, D.S., Ford, V.L.,
Tuohy, I.R., Mills, B.Y., Turtle, A.J., \& Helfand, D.J. 1985, 
ApJS, 58, 197

\bibitem[Magnier et al. (1996)Paper IV]{Paper4} Magnier, E.A, Chu Y.-H.,
Points, S.D., Hwang, U., Smith, R.C. 1996, ApJ, 464, 829 (Paper IV)

\bibitem[Meaburn \& Laspias (1991)]{ML91} Meaburn, J., \& Laspias, N.V.
1991, A\&A, 245, 635

\bibitem[Morrison \& McCammon (1983)]{Morr83} Morrison, R., \& 
McCammon, D. 1983, ApJ, 270, 119

\bibitem[Oey \& Smedley (1998)]{Oey98} Oey, M.S., \& Smedley, S.A. 1998,
AJ, 116, 1263

\bibitem[Osterberg (1997)]{Ost97} Osterberg, J. 1997, private communication

\bibitem[Points et al. (1999)]{Points99} Points, S.D., Chu, Y.-H., Kim, S.,
Smith, R.C., Snowden, S.L., Brandner, W., Greundl, R.A. 1999, ApJ, 518, 298

\bibitem[Pfefferman et al. (1987)]{Pfef87} Pfefferman, E., et al. 1987,
Proc. SPIE, 733, 519

\bibitem[Raymond \& Smith (1977)]{RS77} Raymond, J.C., \& Smith, B.W.
1977, ApJS, 35, 419

\bibitem[Rohlfs et al. (1984)]{Rohlfs84} Rohlfs, K., Kreitschmann, J.,
Feitzinger, J.V., Siegman, B.C. 1984, A\&A, 137, 343

\bibitem[Rosado et al. (1996)]{Rosado96} Rosado, M., Laval, A., 
Le Coarer, E., Georgelin, Y.P., Amram, P., Marcelin, M., Goldes, G., \&
Gach, J.L. 1996, A\&A, 308, 588

\bibitem[{\it ROSAT} Mission Description (1991)]{ROSAT91} {\it ROSAT}
Mission Description, 1991, NASA publication NRA 91-OSSA-25, Appendix F

\bibitem[Sanduleak (1969)]{Sk69} Sanduleak, N. 1969, Cerro Tololo 
Inter-American Obs. Contrib. No. 89

\bibitem[Wang \& Helfand (1991)]{Wang91} Wang, Q., \& Helfand, D.J.
1991, ApJ, 373, 497

\bibitem[Weaver et al. (1977)]{Weaver77} Weaver, R., McCray, R., 
Castor, J., Shapiro, P., \& Moore, R. 1977, ApJ, 208, 610

\bibitem[Williams et al. (1999)]{Williams99} Williams, R.M., Chu, Y.-H.,
Dickel, J.R., Petre, R., Smith, R.C., \& Tavarez, M. 1999, ApJS, 123, 467

\bibitem[Williams et al. (2000)]{Williams00} Williams, R.M., Petre, R.,
Chu, Y.-H., Chen, C.H.R. 2000, ApJ, 536, 27

\bibitem[Willis (1999)]{Willis99} Willis, A. J. 1999, in: K.A. van der Hucht, 
G. Koenigsberger, \& P.R.J. Eenens (eds.), Wolf-Rayet Phenomena in Massive
Star and Starburst Galaxies, Proc. IAU Symp. No. 193 (Chelsea, Mich: Sheridan) 

\end{thebibliography}
\end{document}